See the IEEE published version of this article here: http://dx.doi.org/10.1109/IITC51362.2021.9537515

# Improved Contacts to Synthetic Monolayer MoS$_2$ – A Statistical Study

Aravindh Kumar[1], Alvin Tang[1], H.-S. Philip Wong[1], Krishna Saraswat[1]
[1]Department of Electrical Engineering, Stanford University, Stanford, CA 94305, USA, email: akumar47@stanford.edu

*Abstract*— Two-dimensional (2D) semiconductors are promising candidates for scaled transistors because they are immune to mobility degradation at the monolayer limit. However, sub-10 nm scaling of 2D semiconductors, such as MoS$_2$, is limited by the contact resistance. In this work, we show for the first time a statistical study of Au contacts to chemical vapor deposited monolayer MoS$_2$ using transmission line model (TLM) structures, before and after dielectric encapsulation. We report contact resistance values as low as 330 ohm-um, which is the lowest value reported to date. We further study the effect of Al$_2$O$_3$ encapsulation on variability in contact resistance and other device metrics. Finally, we note some deviations in the TLM model for short-channel devices in the back-gated configuration and discuss possible modifications to improve the model accuracy.

*Keywords—MoS$_2$, contact resistance, 2D semiconductors, transition metal dichalcogenide, transistor*

## I. Introduction

To retain electrostatic control with sub-10 nm gates, the channel body thickness should be scaled below 2 nm [1]. Si mobility degrades significantly at thicknesses below 4 nm, whereas two-dimensional (2D) semiconductors such as transition metal dichalcogenides (TMDs) retain good mobility even at the monolayer limit (< 1 nm thick), making them attractive for use in scaled channels [2]. TMDs such as MoS$_2$ have been extensively studied for use in scaled field-effect transistors because its significant band gap (~1.5 to 2.5 eV) can enable lower leakage power compared to other low-dimensional semiconductors. TMDs can also be directly grown on Si CMOS substrates, without requiring transfer steps like other low-dimensional materials [3], [4], making them attractive for 3D integration.

However, monolayer TMDs are severely limited by contact resistance, characterized by a Schottky-barrier at the metal-semiconductor interface in sub-100 nm channels. This is due to the pinning of the metal Fermi level at the MoS$_2$ interface below the conduction band edge [5]. Au contacts have been shown to make good contacts to 1L MoS$_2$ [2], [6], [7]. However, a systematic statistical study of variation in contact resistance has yet been done. In this work, we study variation in Au contacts to chemical vapor deposited 1L MoS$_2$ and discuss the effects of atomic layer deposited (ALD) Al$_2$O$_3$ encapsulation on contact resistance. We observe that encapsulation reduces the contact resistance as well as its variability. We report the lowest contact resistance ($\approx$ 330 ± 24 Ω.μm) on 1L MoS$_2$ to date and also discuss the accuracy of the TLM model as it is applied to back-gated 2D transistors.

## II. Experimental methods

### A. MoS$_2$ growth

Continuous 1L MoS$_2$ films are grown at 750 °C on 50 nm SiO$_2$ thermally grown in dry O$_2$ on p$^{++}$ Si substrates by solid source chemical vapor deposition (CVD). The growth technique utilizes solid sulfur and molybdenum trioxide precursors with nucleation aided by a perylene-3,4,9,10 tetracarboxylic acid tetrapotassium salt (PTAS) seed layer [3]. The number of MoS$_2$ layers is confirmed by the Raman spectrum shown in Fig. 1 (b).

### B. Device Fabrication

We fabricated TLM structures [8] consisting of back-gated transistors with channel lengths ranging from 100 nm to 1 μm as shown in the scanning electron microscope (SEM) image in Fig 1(a). All patterning steps were done using e-beam lithography. Contact probe pads were first patterned followed by etching away MoS$_2$ in the exposed contact regions using XeF$_2$ vapor. 2 nm Ti / 40 nm Au was deposited by e-beam evaporation at a base pressure of ~10$^{-7}$ Torr. This was followed by overnight liftoff in acetone. Rectangular channels 2 μm wide were then patterned and etched by XeF$_2$. In the same step, the contact pads were isolated from the rest of the MoS$_2$ film by etching away the MoS$_2$ around the pads. Fine contacts were then patterned and 50 nm Au was deposited at a rate of 0.5 A/s by e-beam evaporation at ~10$^{-7}$ Torr base pressure. The metal was lifted off in acetone. Finally, the sample was annealed at 250 °C for 2 hours in vacuum in a Janis vacuum probe station to remove adsorbates from the MoS$_2$ surface [2].

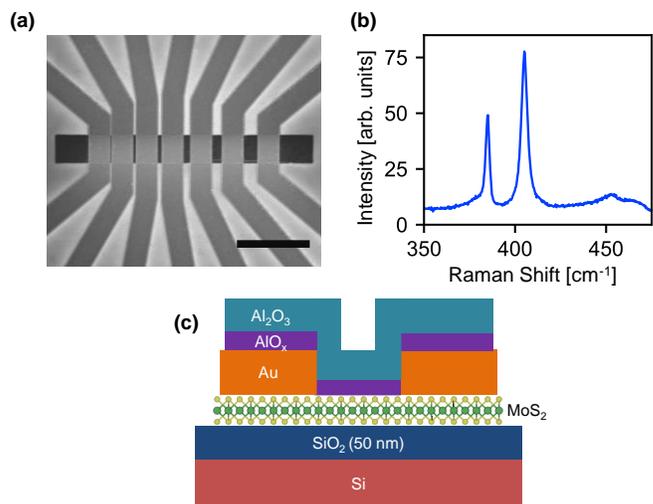

Fig. 1. (a) Scanning electron microscope (SEM) image of a TLM structure with channel lengths ranging from 100 nm to 1 μm. The contact lines are 1.5 μm in width and the MoS$_2$ channel is 2 μm in width. Scale bar is 5 μm. (b) Raman spectrum of 1L MoS$_2$ grown by CVD at 750 °C on 50 nm SiO$_2$ / Si. (c) Schematic cross-section of the final device structure after Al$_2$O$_3$ encapsulation

### C. Dielectric Encapsulation

The chips were encapsulated with atomic layer deposited (ALD) Al$_2$O$_3$. First, a seed layer of 1.5 nm Al was deposited by e-beam evaporation. This layer was oxidized into AlO$_x$ by exposing it to the ambient. 20 nm Al$_2$O$_3$ was subsequently deposited by 200 cycles of ALD at 300 °C after a 300 °C anneal for 30 minutes. The final device structure is shown in Fig. 1(c).

## D. Electrical Testing

Electrical testing was carried out in a Cascade Microtech Summit semi-automated probe station with an Agilent B1600A parameter analyzer. After loading the chip, the probe station was purged with $N_2$ for 30 minutes before starting the measurements to ensure the removal of moisture from the $MoS_2$ surface. The chip contains 72 TLM structures, each containing 6 channel lengths – 100 nm, 200 nm, 300 nm, 500 nm, 700 nm and 1 µm – a total of 432 transistors. The devices were tested before and after the encapsulation with $Al_2O_3$.

## III. RESULTS AND DISCUSSION

Fig. 2 shows the drain current ($I_D$) vs gate-to-source voltage ($V_{GS}$) transfer characteristics measured before and after the $Al_2O_3$ encapsulation for each channel length. The encapsulation results in a threshold voltage ($V_T$) shift, indicating that the $Al_2O_3$ layer causes mild n-type doping in the channel. Fig. 3(b) also shows the $V_T$ shift in 100 nm channels due to encapsulation. Previous works have shown that defective ALD $Al_2O_3$ deposited at low-temperatures cause n-type doping in sulfur-based TMDs [6], [9], [10].

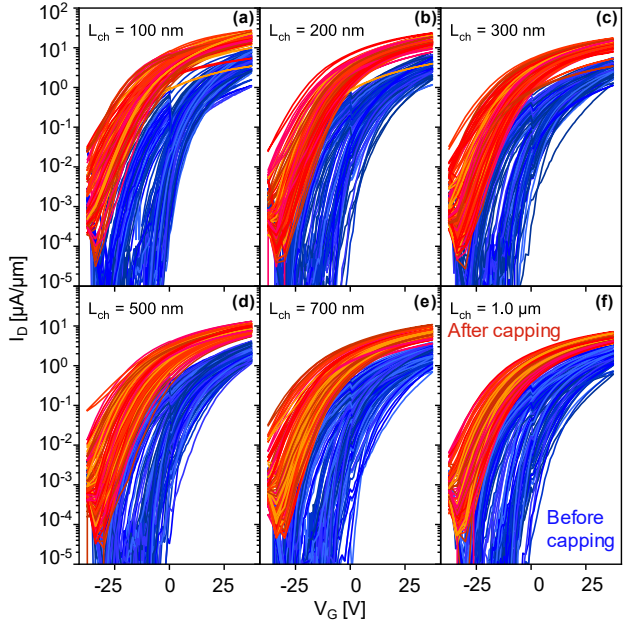

Fig. 2. Measured transfer characteristics ($I_D$-$V_{GS}$) for channel lengths (a) 100 nm, (b) 200 nm, (c) 300 nm, (d) 500 nm, (e) 700 nm and (d) 1 µm; blue (red) curves are measured before (after) $Al_2O_3$ encapsulation. All measurements are done with a $V_{DS}$ of 0.1 V at room temperature with $N_2$ purging.

Because $Al_2O_3$ is used here for purposes of encapsulation only, $Al_2O_3$ was deposited at a higher temperature of 300 °C which leads to a relatively less defective film, therefore reducing the doping effect. Significant on-off ratio is retained as well without any post-encapsulation anneal [6].

For each TLM structure, the contact resistance was extracted from the $I_D$-$V_{GS}$ characteristics by first aligning the threshold voltages of all the channels in that TLM structure and then carrying out linear regression of the total resistance versus channel length. The $MoS_2$ film shows spatial variation across the chip in grain size, film coverage and proportion of bilayer island regions [11]. Therefore, it is important to have each TLM structure spatially localized to minimize the variation across its channel lengths. Fig. 3 shows the cumulative density function (CDF) plot of the extracted contact resistances for all TLM structures, extracted at a carrier density $n = 8\times10^{12}$ cm$^{-2}$. We discarded TLM fits with R-squared less than 0.9 while relatively few TLM structures had R-squared > 0.9 before encapsulation. The median contact resistance is 3.2 kΩ.µm.

The CDF plot shows that encapsulation reduces the average contact resistance as well as the variation across the chip when compared at the same carrier density level. This could be due to two reasons. First, encapsulation improves the sheet resistance at the same carrier density due to improvement in the effective mobility as shown in Fig. 3(c). This improvement in mobility is due to effective screening of charged impurities in $MoS_2$ by the high dielectric constant $Al_2O_3$ layer [12], [13]. This leads to a larger proportion of the drain-to-source voltage dropped across the Schottky contacts, which leads to a lower contact resistance. Second, the Schottky barrier height might be reduced due to induced strain in the $MoS_2$ due to built-in tensile stress from the encapsulation layer. [14], [15], [16]

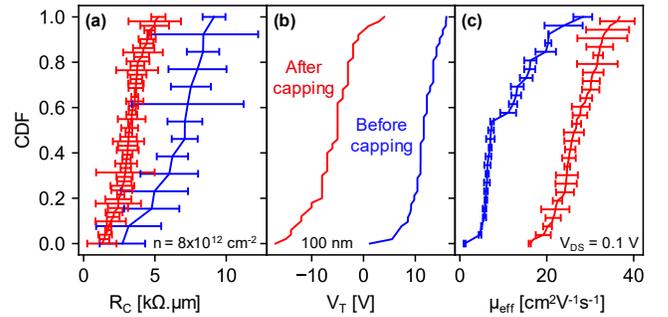

Fig. 3. Cumulative density function (CDF) plot for various device metrics before (blue) and after (red) encapsulation. (a) CDF plot of contact resistance values extracted from TLM structures at carrier density $n = 8\times10^{12}$ cm$^{-2}$. Each datapoint in the curve corresponds to contact resistance extracted from a single TLM structure and the horizontal error bars are obtained from the standard error of the respective linear regression parameters. (b) CDF plot of threshold voltages ($V_T$) in 100 nm channels shows moderate doping due to $Al_2O_3$ layer. (c) CDF plot of effective mobility at carrier density $n = 8\times10^{12}$ cm$^{-2}$.

The lowest extracted contact resistance is 330 ± 24 Ω.µm at carrier density $n = 1.4\times10^{13}$ cm$^{-2}$, which is, to the best of our knowledge, the lowest reported to date on 1L $MoS_2$. Fig. 4 shows the electrical characteristics of the corresponding TLM structure. Linear regression results in an excellent TLM fit with R-squared ≈ 1.0. Further, we have benchmarked the best result from this work with other reported values in literature in Fig. 5(a). This work shows an improvement over previous work on $Al_2O_3$ encapsulated 1-L $MoS_2$ [6] due to the $MoS_2$ being grown at a higher temperature of 750 °C which results in a larger density of bilayer regions [11]. These bilayer regions reduce the Schottky barrier height due to the smaller bandgap of bilayer $MoS_2$ compared to 1L $MoS_2$. Another factor is the use of a thicker $Al_2O_3$ film. The built-in stress in ALD $Al_2O_3$ increases with thickness which could induce greater tensile strain in $MoS_2$, thus reducing the Schottky barrier height [15].

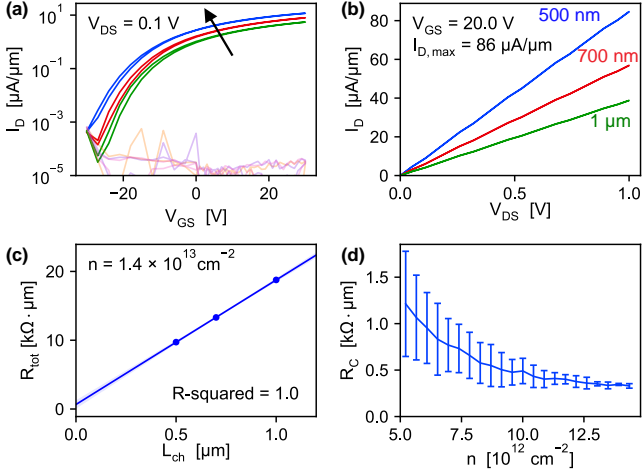

Fig. 4. Electrical characteristics of the TLM structure with the lowest contact resistance: (a) $I_D$ vs. $V_{GS}$ characteristics measured for various channel lengths at $V_{DS}$ = 0.1 V. Red arrow shows the direction of decreasing channel lengths – 1 µm, 700 nm and 500 nm. Gate current is plotted as faded translucent curves. (b) $I_D$ vs. $V_{DS}$ characteristics measured for various channel lengths at $V_{GS}$ = 20 V. (c) Results of the linear regression fit at n = 1.4x10$^{13}$ cm$^{-2}$, with R-squared = 0.99999. (d) Contact resistance ($R_C$) vs carrier density (n) at $V_{DS}$ = 0.1 V. The y-intercept of the fit line is equal to 2$R_C$. The lowest contact resistance value obtained is 330 ± 24 Ω.µm at n = 1.4x10$^{13}$ cm$^{-2}$.

Fig. 5(b) shows a plot of total resistance versus channel length for all TLM structures combined. Although this analysis masks the effects of spatial variation in MoS$_2$ across the chip, especially when compared to Fig. 3(a), it provides an averaged conservative estimate for the contact resistance.

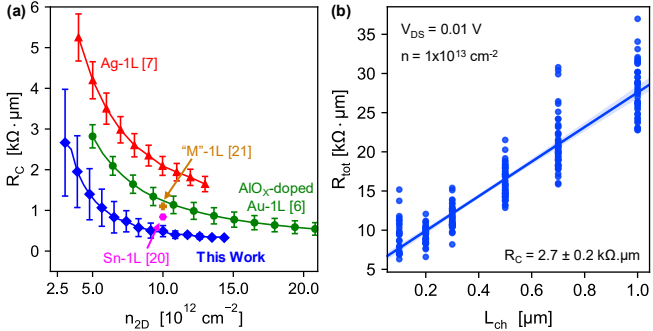

Fig. 5. (a) Benchmark plot comparing the contact resistance ($R_C$) versus carrier density results after encapsulation with other published works [6], [7], [20], [21]. We achieve a low value of $R_C$ = 330 ± 24 Ω.µm at n = 1.4x10$^{13}$ cm$^{-2}$ (adapted from [6]). (b) Total resistance plotted versus channel length for all TLM structures at $V_{DS}$ = 0.01 V. Resistance at 100 nm channel length clearly deviates from the linear trend. Ordinary least-squares linear regression results: $R_C$ = 2.75 ± 0.16 kΩ.µm.

Fig. 5(b) also reveals apparent deficiencies in the TLM model as it is used currently. The total resistance values for the 100 nm channel length deviate from the linear trend followed by the longer channel lengths. This could be explained by the effect of contact-gating on $V_T$ extraction [17], [18]. Contact resistance forms a larger proportion of the total resistance in shorter channels and thus, contact-gating influences $V_T$ extraction more strongly. This could influence the process of adjusting the total resistance of the various channels using their threshold voltages before linear fitting. The fringing electric field of the back-gate capacitor could also explain this deviation since the effect on gate capacitance is significant for shorter channels [19].

## IV. CONCLUSIONS

We have studied Au contacts on 1L MoS$_2$ and the effects of ALD encapsulation using Al$_2$O$_3$. We have demonstrated contact resistance as low as 330 ± 24 Ω.µm which is the lowest reported to date for 1L MoS$_2$. By testing a large array of TLM structures, we have gained statistical insight into the contact resistance variation. The average contact resistance is improved by encapsulation at the same carrier density level. This could be due to the improved effective mobility with encapsulation which leads to a larger voltage drop across the Schottky contacts. Another possible reason is the built-in stress in ALD Al$_2$O$_3$ encapsulation which could induce strain in MoS$_2$, thereby reducing the Schottky barrier height. Encapsulation also reduces variation in contact resistance as well as effective mobility. We note that 100 nm channel devices deviate from the linear TLM trend of total resistance. The TLM model requires modifications to account for the fringing electric field in the back gate capacitor and the effects of contact-gating on extracted $V_T$, These effects become quite significant in short channel devices.

## ACKNOWLEDGEMENTS

Fabrication, material characterization were performed at the Stanford Nanofabrication Facility (SNF) and the Stanford Nano Shared Facilities (SNSF). This work was supported in part by the Stanford SystemX Alliance and Samsung Semiconductor.

## REFERENCES

[1] T. Skotnicki, J. A. Hutchby, T. J. King, H. S. P. Wong, and F. Boeuf, *IEEE Circuits Devices Mag.*, vol. 21, no. 1, pp. 16–26, 2005.
[2] C. D. English, G. Shine, V. E. Dorgan, K. C. Saraswat, and E. Pop, *Nano Lett.*, vol. 16, no. 6, pp. 3824–3830, 2016.
[3] K. K. H. Smithe, C. D. English, S. V. Suryavanshi, and E. Pop, *2D Mater.*, vol. 4, no. 1, pp. 1–40, 2017.
[4] I. Asselberghs *et al. IEDM Tech Dig*, pp. 893–896, 2020.
[5] S. Das, H. Y. Chen, A. V. Penumatcha, and J. Appenzeller, *Nano Lett.*, vol. 13, no. 1, pp. 100–105, 2013.
[6] C. J. McClellan, E. Yalon, K. K. H. Smithe, S. V. Suryavanshi, and E. Pop, *ACS Nano*, vol. 15, no. 1, pp. 1587–1596, Jan. 2021.
[7] K. K. H. Smithe, C. D. English, S. V. Suryavanshi, and E. Pop, *Nano Lett.*, vol. 18, no. 7, pp. 4516–4522, 2018.
[8] H. H. Berger, *Solid State Electron.*, vol. 15, no. 2, pp. 145–158, 1972.
[9] A. Leonhardt *et al.*, *ACS Appl. Mater. Interfaces*, vol. 11, no. 45, pp. 42697–42707, 2019.
[10] A. Kumar, K. N. Nazif, P. Ramesh, and K. Saraswat, *Device Res. Conf. - Conf. Dig. DRC*, vol. 2020-June, no. 11, pp. 4–5, 2020.
[11] K. K. H. Smithe, S. V. Suryavanshi, M. Muñoz Rojo, A. D. Tedjarati, and E. Pop, *ACS Nano*, vol. 11, no. 8, pp. 8456–8463, 2017.
[12] D. Jena and A. Konar, *Phys. Rev. Lett.*, vol. 98, no. 13, p. 136805, 2007.
[13] Z. Yu *et al.*, *Adv. Mater.*, vol. 28, no. 3, pp. 547–552, 2016.
[14] G. Krautheim, T. Hecht, S. Jakschik, U. Schröder, and W. Zahn, *Appl. Surf. Sci.*, vol. 252, no. 1, pp. 200–204, Sep. 2005.
[15] A. P. John, A. Thenapparambil, and M. Thalakulam, *Nanotechnology*, vol. 31, no. 27, p. 275703, 2020.
[16] J. Quereda, J. J. Palacios, N. Agräit, A. Castellanos-Gomez, and G. Rubio-Bollinger, *2D Mater.*, vol. 4, no. 2, p. 021006, Jan. 2017.
[17] J. R. Nasr, D. S. Schulman, A. Sebastian, M. W. Horn, and S. Das, *Adv. Mater.*, vol. 31, no. 2, pp. 1–9, 2019.
[18] A. Prakash, H. Ilatikhameneh, P. Wu, and J. Appenzeller, *Sci. Rep.*, vol. 7, no. 1, pp. 1–9, 2017.
[19] S. Heinze, M. Radosavljević, J. Tersoff, and P. Avouris, *Phys. Rev. B - Condens. Matter Mater. Phys.*, vol. 68, no. 23, pp. 1–5, 2003.
[20] A.-S. Chou *et al.*, *IEEE Electron Device Lett.*, vol. 42, no. 2, pp. 272–275, 2020.
[21] A.-S. Chou *et al.*, *2020 IEEE Symposium on VLSI Technology*, Jun. 2020, pp. 1–2.